\documentclass[12pt,a4paper]{article}
\usepackage{amssymb}
\usepackage[unicode]{hyperref}
\begin{document}
\def\emline#1#2#3#4#5#6{%
       \put(#1,#2){\special{em:moveto}}%
       \put(#4,#5){\special{em:lineto}}}
\def\newpic#1{}
\title{The pedagogical value of the four-dimensional picture~III:
Solutions to Maxwell's equations}
\author{Andrew E Chubykalo${}^\dagger$, Augusto Espinoza${}^\dagger$
and B P Kosyakov${}^{\ddagger \S}$}
\maketitle
\begin{center}
{\small${}^\dagger$ 
Escuela de F{\'i}sica, 
Universidad Aut\'onoma de Zacatecas, 
Apartado Postal C-580 Zacatecas 98068, Zacatecas,  
Mexico\\
${}^\ddagger$ Russian Federal Nuclear Center, 
Sarov, 607189 Nizhni\u{\i} Novgorod Region, Russia \\
${}^\S$ Moscow Institute of Physics {\&} Technology, Dolgoprudni\u{\i}, 141700 Moscow Region, 
Russia\\
} 
\end{center}

\begin{abstract}
{\noindent
We outline a regular way for solving Maxwell's equations.
We take, as the starting point, the notion of vector potentials. 
The rationale for introducing this notion in electrodynamics is 
that the set of Maxwell's equations is seemingly overdetermined.
We demonstrate the existence of two fundamental solutions to Maxwell's equations 
whose linear combinations comprise the whole variety of classical 
electromagnetic field configurations.}
\end{abstract}


\tableofcontents

\section{Introduction}
\label
{Introduction}

In this, third, paper of a series of papers, initiated by Refs.~\cite{4D-1, 4D-2}, 
we continue to review the utility of four-dimensional concepts 
in classical electrodynamics. 

The discussion of Ref.~\cite{4D-2} made its clear that the law governing the 
electromagnetic field behavior is largely ordered by the geometry of Minkowski 
spacetime ${\mathbb R}_{1,3}$.
This law is given by a system of partial differential equations 
\begin{equation}
\partial_\lambda F^{\lambda\mu}=4\pi j^\mu\,,
\label
{Maxw-second}
\end{equation}                                          
\begin{equation}
\partial_\lambda{}^\ast\!F^{\lambda\mu}=0\,,
\label
{Maxw-first}
\end{equation}                                          
known as Maxwell's equations.
Here $\partial_\lambda$ stands for $\partial/\partial_\lambda$, and 
${}^\ast\!F^{\lambda\mu}\equiv\frac12\,\epsilon^{\lambda\mu\nu\rho}F_{\nu\rho}$.
We use the Gaussian system of units, 
and put the speed of light to be 1.
With fixing a particular inertial frame of reference, Eqs.~(\ref{Maxw-second}) 
and (\ref{Maxw-first}) can be rewritten as  
\begin{equation}
{\nabla}\cdot{\bf E}=4\pi\varrho\,,
\label
{Maxw-div-E}
\end{equation}                                          
\begin{equation}
{\nabla}\times{\bf B}=4\pi{\bf j}+\frac{\partial{\bf E}}{\partial t}\,, 
\label
{Maxw-rot-B}
\end{equation}                                          
\begin{equation}
{\nabla}\cdot{\bf B}=0\,,
\label
{Maxw-div-B}
\end{equation}                                          
\begin{equation}
{\nabla}\times{\bf E}=-\frac{\partial{\bf B}}{\partial t}\,. 
\label
{Maxw-rot-E}
\end{equation}                                          
Although the set of differential equations in four-dimensional tensor form, 
Eqs.~(\ref{Maxw-second})--(\ref{Maxw-first}), is mathematically equivalent to 
the set of differential equations in three-dimensional vector form, 
Eqs.~(\ref{Maxw-div-E})--(\ref{Maxw-rot-E}), 
the former is much more elegant than the latter.

We now address the issue of whether the four-dimensional covariant treatment of 
basic solutions to Maxwell's equations is favored over the corresponding 
three-dimensional vector treatment which is adopted in modern textbooks on 
electrodynamics and used as a common practice for teaching these solutions at 
undergraduate level.
The aim of this paper is to show that this is indeed the case.
Our main concern is with two questions: 

(i) Where did the notion of vector potentials come from?

(ii) Are there several fundamental solutions to Maxwell's equations such that 
their linear combinations form the whole variety of classical electromagnetic 
fields distributed over empty space? 

The rationale, or at least a motivation for introducing vector potentials is 
that the set of Maxwell's equations is seemingly overdetermined.
A regular procedure for solving this set of differential 
equations is to express the  electromagnetic field strength $F_{\mu\nu}$ in 
terms of vector potentials $A_\mu$.

As to the second question, the answer is positive.
In fact, there exist two fundamental solutions to Maxwell's equations
whereby every electromagnetic field configuration can be constructed.
We will see that  the {Li\'enard--Wiechert} field and {plane wave} are
acting as fundamental solutions of this kind in classical electrodynamics.

It will transpire that the four-dimensional covariant framework not only makes 
the analysis of the posed questions much easier, but also provides a
decisive pedagogical insight into geometric and physical 
information which is encoded in Maxwell's equations.  

\section{The notion of vector potentials}  
\label
{gauge fields}

At first glance the set of Maxwell's equations, Eqs.~(\ref{Maxw-second})--(\ref{Maxw-first}), 
is {\it overdetermined}: 8 equations are intended for finding 6 unknown 
functions $F^{\mu\nu}$.
Matters can be improved by expressing the field strength $F^{\mu\nu}$ in terms of 
vector potentials $A^{\mu}$, 
\begin{equation}
F^{\mu\nu}=\partial^\mu A^{\nu}-\partial^\nu A^{\mu}\,.
\label
{F-in-terms-A}
\end{equation}                                          
Recall that  $F^{\mu\nu}$ is an antisymmetric tensor, so that the antisymmetric 
combination of two vectors $\partial^\mu$ and $A^\nu$ on the right of 
(\ref{F-in-terms-A}) is quite appropriate. 
With the {ansatz} (\ref{F-in-terms-A}), the second part of Maxwell's equation,
Eq.~(\ref{Maxw-first}), is satisfied {identically} because
$\epsilon^{\lambda\mu\nu\rho}\partial_\lambda\partial_\nu\equiv 0$.
Substituting (\ref{F-in-terms-A}) into (\ref{Maxw-second}) gives 
\begin{equation}
\Box A^{\mu}-\partial^\mu \partial_\lambda A^\lambda=4\pi{j}^\mu\,\,,
\label
{Maxw-A-j} 
\end{equation}                                          
where $\Box=\partial^\lambda \partial_\lambda={\partial^2}/{\partial t^2}-\nabla^2$ 
is the {wave operator}. 
We thus come to the set of equations, Eq.~(\ref{Maxw-A-j}), 
which has the number of equations equal to the number of the functions sought.

Note that $A^{\mu}$ is defined in Eq.~(\ref{F-in-terms-A}) up to adding the 
four-gradient of an arbitrary smooth scalar function $\partial^\mu\chi$. 
Indeed, the field strength $F^{\mu\nu}$ is unaffected by the replacement 
\begin{equation}
{A}^\mu\to{A'}^{\!\mu}={A}^\mu-\partial^\mu\chi\,.
\label
{gauge-trans-4D}
\end{equation}                                          
These transformations of $A^{\mu}$ are called  {\it gauge transformations}. 
We thus deal with the entire equivalence class of vector potentials related to each 
other by gauge transformations, rather than a concrete vector function. 
The term  $\partial^\mu\chi$ in (\ref{gauge-trans-4D})
is called the {\it gauge mode}.
These modes do not contribute to the Lorentz force $qv^\nu F_{\mu\nu}$, and hence 
the dynamics of charged particles is unaffected by them.
On the other hand, the current of charged particles  $j^\mu$ is not the source 
of gauge modes. 
Indeed, it is clear that gauge modes satisfy Eq.~(\ref{Maxw-A-j}) with the 
vanishing right side of this equation, whence it follows that gauge modes are 
unaffected by $j^\mu$,
and their evolution is divorced from the evolution of the dynamical 
degrees of freedom described by $F_{\mu\nu}$. 

This offers a clearer view of how the seemingly overdetermined set of partial differential 
equations becomes determined. 
The net dynamical degrees of freedom are 
augmented by the addition of auxiliary degrees of freedom, gauge  modes, 
which equalizes the number of equations governing this extended field system to 
the number of field variables.

The corresponding treatment of Maxwell's equations in three-dimensional vector 
form, Eqs.~(\ref{Maxw-div-E})--(\ref{Maxw-rot-E}), is not as much intelligible. 
Let us write  components of ${A}^\mu$  in a particular inertial frame: 
${A}^\mu=(\phi,{\bf A})$, or, equivalently,  ${A}_\mu=(\phi,-{\bf A})$.
Taking into account the definitions of the electric field $E_i=F_{0i}$ and the 
magnetic induction $B_i=-{\scriptstyle\frac12}\,\epsilon_{ijk}F^{jk}$ which
were given in \cite{4D-2} we obtain from (\ref{F-in-terms-A}) 
\begin{equation}
{\bf E}=-\frac{\partial{\bf A}}{\partial t}-\nabla\phi\,,
\label
{E=nabla-phi}
\end{equation}                                          
\begin{equation}
{\bf B}=\nabla\times{\bf A}\,.
\label
{B=-rot-A-}
\end{equation}                                          
It is then possible to verify, by inspection, that Eqs.~(\ref{E=nabla-phi}) and 
(\ref{B=-rot-A-}) provide a solution of equations (\ref{Maxw-div-B}) and 
(\ref{Maxw-rot-E}).
However, if the four-dimensional ansatz (\ref{F-in-terms-A}) was not taken as 
the starting point, 
then the three-dimensional ansatz (\ref{E=nabla-phi})--(\ref{B=-rot-A-}) 
is an ingenious mathematical trick whose discovery is surprising. 

The corresponding three-dimensional gauge transformations are 
\begin{equation}
\phi\to\phi'=\phi-\frac{\partial \chi}{\partial t}\,,
\label
{gauge-trans-3D-phi}
\end{equation}                                          
\begin{equation}
 {\bf A}\to{\bf A}'={\bf A}+\nabla \chi\,\,.
\label
{gauge-trans-3D}
\end{equation}                                          

It is unlikely that Eqs.~(\ref{E=nabla-phi})--(\ref{B=-rot-A-}) and 
(\ref{gauge-trans-3D-phi})--(\ref{gauge-trans-3D}) might help 
to illuminate the origin and mathematical nature of  $\phi$ and 
${\bf A}$.
{Historically, the ansatz (\ref{B=-rot-A-}) was suggested by W. Thomson 
who investigated the analogies of electric phenomena with those of elasticity, 
and by C. Neumann, Weber and Kirchhoff in their studies on the induction 
of currents \cite{Whittaker}.}
However, while on the subject of modern teaching in gauge field theory
(specifically in electrodynamics) following its
inner logic, and not according to its historical development,
it is apparent that the student should learn of the notion of vector potentials 
in the four-dimensional relativistic framework. 

Equation (\ref{Maxw-first}) can be rearranged to give\footnote{To see this, let us  
note that $\partial_\lambda{}^\ast\!F^{\lambda\rho}=\frac12\,\epsilon^{\lambda\rho\mu\nu}\partial_\lambda F_{\mu\nu}$
is proportional to the sum of terms stemming from the antisymmetrization of 
$\partial_\lambda F_{\mu\nu}$.
Among them, the plus sign have terms which can be represented as 
cyclic permutations of indices of $\partial_\lambda F_{\mu\nu}$,
that is, $\partial_\lambda F_{\mu\nu}$, $\partial_\nu F_{\lambda\mu}$ and $\partial_\mu F_{\nu\lambda}$, 
while the terms of another triplet
$\partial_\lambda F_{\nu\mu}$, $\partial_\mu F_{\lambda\nu}$ and $\partial_\nu F_{\mu\lambda}$
are assigned the minus sign.
However, both triplets actually contain identical terms because
$F_{\mu\nu}=-F_{\nu\mu}$.
It follows that equations  (\ref{Maxw-first}) and (\ref{Maxw-tens-first})
are equivalent.
}
\begin{equation}
\partial_\lambda F_{\mu\nu}+
\partial_\nu F_{\lambda\mu}+\partial_\mu F_{\nu\lambda}=0\,.
\label
{Maxw-tens-first}
\end{equation}                                          
We call equations (\ref{Maxw-first}) and (\ref{Maxw-tens-first}) collectively 
the {\it Bianchi identity}.
This name comes from the fact that if we adopt $A_\mu$ as the basic variables,
then the possibility to express $F_{\mu\nu}$ in terms of  $A_\mu$, as shown in 
Eq.~(\ref{F-in-terms-A}), becomes a synonym for making the left sides of equations 
(\ref{Maxw-first}) and (\ref{Maxw-tens-first}) identitically vanishing.
                        
Equation (\ref{Maxw-A-j}) cannot be solved {\it directly} because
the differential operator $\Lambda^{\mu}_{~\lambda}(\partial)=
\delta^{\mu}_{~\lambda}\,\Box-\partial^\mu\partial_\lambda$
has no inverse\footnote{The expert reader will recognize that 
${\rm det}\,{\Lambda}=\lambda_0\lambda_1\lambda_2\lambda_3$
where $\lambda_0$, $\lambda_1$, $\lambda_2$, $\lambda_3$ are eigenvalues of
the operator $\Lambda^{\mu}_{~\lambda}(k)=k^2\delta^{\mu}_{~\lambda}-
k^{\mu}k_{\lambda}$, that is, solutions to the eigenvalue problem
\begin{equation}
\Lambda^{\mu}_{~\lambda}\Psi^\lambda_a=\lambda_a\Psi^\mu_a\,,
\label
{Lambda-Sec4.2}
\end{equation}                                          
where $a$ runs from 0 to 3 (there is no summation over $a$ in the right-hand 
side).
It is clear from 
\begin{equation}
\Lambda^{\mu}_{~\lambda}(k)k^{\lambda}=0
\label
{Lambda-k-0}
\end{equation}                                          
that some eigenvalue (associated with the eigenvector $\Psi^\mu$ proportional 
to $k^\mu$) is zero, hence ${\rm det}\,{\Lambda}=0$.}.
To tackle this problem, we take advantage of the gauge arbitrariness.
We impose a gauge fixing condition on ${A}_{\mu}$, selecting a single representative of the equivalence class of vector potentials, to
make the differential operator  
invertible.
For example, if we choose the so-called Lorenz gauge condition
\begin{equation}
\partial_\lambda {A}^\lambda=0\,, 
\label
{gauge}
\end{equation}  
then (\ref{Maxw-A-j}) becomes the inhomogeneous wave equation
\begin{equation}
\Box {A}^{\mu}=4\pi{j}^\mu\,\,.
\label
{wave-A-j}
\end{equation}  
This resolves the problem because the wave operator $\Box$ is invertible.   

\section{Fundamental solutions to Maxwell's equations}
\label
{solutions}
It would be difficult if not impossible to find in current texts on classical 
electrodynamics the statement that all feasible classical electromagnetic field 
configurations can proliferate through composing linear combinations of only 
{\it two} fundamental solutions to Maxwell's equations\footnote{To illustrate we 
refer to 
several popular textbooks \cite{Jackson, LandauLifshitz, Synge, Barut, 
Rohrlich5} where this fact went unnoticed.}.
We intend to show that two electromagnetic field configurations, known as
the Lien\'ard--Wiechert field and {plane wave},
may be regarded as fundamental solutions of this kind. 

It is sufficient to restrict our consideration to  
${A}_{\mu}$  generated by a single charged particle moving along a smooth 
timelike world line $z^\mu(s)$.                                   
We can readily extend this analysis to cover the case of several charged particles 
by taking the the sum of all such ${A}_{\mu}$'s generated by their respective 
individual sources.

Thus, our concern is with finding exact solutions to the equation
\begin{equation}
\Box {A}^{\mu}(x)=4\pi q\int_{-\infty}^\infty ds\,{v}^\mu(s)\delta^4[x-z(s)]\,,
\label
{wave-A-single}
\end{equation}  
in which ${A}^{\mu}(x)$ is the unknown variable, $e$ is the charge of the particle, 
and $v^\mu=dz^\mu/ds$ its four-velocity.
Since the partial differential equation (\ref{wave-A-single}) is linear in 
 ${A}^{\mu}(x)$, its solution is written as the sum of some particular solution 
of this equation and general solution of the associated homogeneous equation 
\begin{equation}
\Box {A}^{\mu}=0\,.
\label
{wave-A}
\end{equation}   

\begin{figure}[htb]
\begin{center}
\unitlength=1.00mm
\special{em:linewidth 0.4pt}
\linethickness{0.4pt}
\begin{picture}(34.00,35.00)
\put(10.00,10.00){\vector(0,1){10.00}}
\put(10.00,10.00){\vector(1,1){20.00}}
\emline{10.00}{30.00}{1}{12.00}{30.00}{2}
\emline{13.00}{30.00}{3}{15.00}{30.00}{4}
\emline{16.00}{30.00}{5}{18.00}{30.00}{6}
\emline{19.00}{30.00}{7}{21.00}{30.00}{8}
\emline{22.00}{30.00}{9}{24.00}{30.00}{10}
\emline{25.00}{30.00}{11}{27.00}{30.00}{12}
\emline{28.00}{30.00}{13}{30.00}{30.00}{14}
\emline{30.00}{10.00}{15}{30.00}{12.00}{16}
\emline{30.00}{13.00}{17}{30.00}{15.00}{18}
\emline{30.00}{16.00}{19}{30.00}{16.00}{20}
\emline{30.00}{16.00}{21}{30.00}{18.00}{22}
\emline{30.00}{19.00}{23}{30.00}{21.00}{24}
\emline{30.00}{22.00}{25}{30.00}{24.00}{26}
\emline{30.00}{25.00}{27}{30.00}{27.00}{28}
\emline{30.00}{28.00}{29}{30.00}{30.00}{30}
\put(2.00,9.00){\makebox(0,0)[cc]{$z^\mu(s_{\rm ret})$}}
\put(7.00,19.00){\makebox(0,0)[cc]{$v^\mu$}}
\put(8.00,30.00){\makebox(0,0)[cc]{$\rho$}}
\put(30.00,8.00){\makebox(0,0)[cc]{$\rho$}}
\put(34.00,33.00){\makebox(0,0)[cc]{$x^\mu$}}
\put(22.00,18.00){\makebox(0,0)[cc]{$R^\mu$}}
\bezier{44}(11.00,5.00)(9.00,10.00)(11.00,15.00)
\bezier{64}(11.00,15.00)(15.00,23.00)(13.00,30.00)
\bezier{24}(13.00,30.00)(11.00,35.00)(11.00,37.00)
\bezier{56}(20.00,20.00)(25.00,25.00)(30.00,30.00)
\bezier{44}(10.00,10.00)(15.00,10.00)(20.00,10.00)
\bezier{44}(20.00,10.00)(25.00,10.00)(31.00,10.00)
\bezier{44}(10.00,20.00)(10.00,25.00)(10.00,31.00)
\end{picture}
\caption{The origin of the invariant retarded variable $\rho$}
\label
{covariant}
\end{center}
\end{figure}
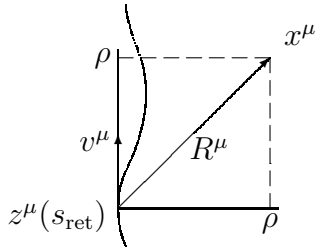

What is the most appropriate form of the particular solution to equation 
(\ref{wave-A-single}) for the description of the classical electromagnetic 
picture?
The commonly accepted point of view is that the {\it retarded} vector potential
${{A}}^{\mu}_{\rm ret}$, called the Li\'enard--Wiechert potential, is just
this solution. 
The procedure of derivation of the Li\'enard--Wiechert potential is outlined in every textbook.
We thus only recall the form of this solution using the condensed 
four-dimensional Dirac notations \cite{Dirac}. 
Let $x^\mu$ be some point outside the world line $z^\mu(s)$. 
Define the lightlike four-vector  $R^\mu=x^\mu-z^\mu(s_{\rm ret})$ 
drawn from a point $z^\mu(s_{\rm ret})$ on the world line, where the signal
is emitted, to the point $x^\mu$, where the signal is received.
\begin{figure}[htb]
\begin{center}
\unitlength=1mm
\special{em:linewidth 0.4pt}
\linethickness{0.4pt}
\begin{picture}(45.00,42.00)
\emline{15.00}{15.00}{1}{30.00}{30.00}{2}
\emline{30.00}{30.00}{2}{40.00}{20.00}{3}
\emline{40.00}{20.00}{3}{45.00}{15.00}{4}
\bezier{128}(45.00,15.00)(30.00,9.00)(15.00,15.00)
\bezier{52}(21.00,18.00)(19.00,24.00)(21.00,30.00)
\bezier{52}(21.00,30.00)(24.00,37.00)(23.00,42.00)
\bezier{20}(24.00,12.00)(25.00,10.00)(25.00,7.00)
\put(30.00,30.00){\makebox(0,0)[cc]{$\bullet$}}
\put(32.00,34.00){\makebox(0,0)[cc]{$x^\mu$}}
\put(17.00,40.00){\makebox(0,0)[cc]{$z^\mu(s)$}}
\put(28.00,18.00){\makebox(0,0)[cc]{$z^\mu(s_{\rm ret})$}}
\end{picture}
\caption{Retarded signal at $x^\mu$, emitted from $z^\mu(s_{\rm ret})$}
\label
{past-cone}
\end{center}
\end{figure}
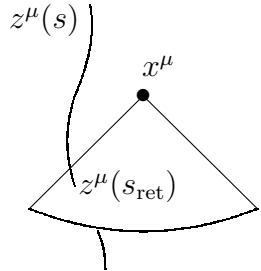
It is seen from Figure \ref{past-cone} that $R^\mu$ is opposed to 
a ray of the past light cone with the vertex at $x^\mu$.
Consider the unit vector  $v^\mu$ tangent
to the curve  $z^\mu(s)$ at the instant $s_{\rm ret}$ and define the scalar
\begin{equation}
\rho=R_\alpha v^\alpha\,.
\label
{rho-def}
\end{equation}                                          
Since $R^\mu$ is a lightlike vector, the geometric interpretation of
$\rho$ is apparent: $\rho$ is the spatial distance between the
field point and the retarded point in the instantaneously comoving Lorentz
frame in which the charge is at rest at the retarded instant $s_{\rm ret}$, as
shown in Figure~\ref{covariant}.

The retarded  vector potential due to a single 
arbitrarily moving charge $q$ is
\begin{equation}
{{A}}^{\mu}_{\rm ret}(x)=q\,\frac{v^\mu}{\rho}\,.
\label
{L-W-potential}
\end{equation}  

Note that this expression for the retarded vector potential can be directly 
derived from that for the Coulomb potential, with the understanding that 
the retardation condition is met, see, e. g., Ref.~\cite{LandauLifshitz}.

The strength of the Li\'enard--Wiechert field is readily calculated from 
(\ref{L-W-potential}) with the aid of simple differentiation rules 
(these rules can be found in the textbooks \cite{Synge}, 
\cite{Barut}, \cite{Rohrlich5}).
The result is\footnote{A diligent student will perform this calculation as a 
useful exercise.}  
\begin{equation}
F_{~{\rm ret}}^{\mu\nu}=q\,\frac{R^\mu U^\nu-R^\nu U^\mu}{\rho^3}\,, 
\label
{F-L-W}
\end{equation}                       
\begin{equation}
U^\mu=\left(1-a_\alpha R^\alpha\right){v^\mu}+\rho\,{a^\mu}\,,
\label
{V}
\end{equation}                       
where $a^\mu=dv^\mu/ds$ is the four-acceleration of the charged particle.
 
It might be well to point out that the four-dimensional description 
represents the Li\'enard--Wiechert field, Eqs.~(\ref{F-L-W})--(\ref{V}), in a 
concise and elegant form.
In contrast, the conventional three-dimensional vector treatment leads 
to 
rather cumbersome expressions
\begin{equation}
{\bf E}_{\rm ret}=\frac{q}{\left(r-{\bf r}\cdot{\bf v}\right)^3}\left\{{\left(1-{\bf v}^2\right)
\left({\bf r}-r{\bf v}\right)}
+
{{\bf r}\times\left[\left({\bf r}-r{\bf v}\right)\times{\bf a}\right] 
}
\right\}, 
\label
{E-L-W}
\end{equation}                       
\begin{equation}
{\bf B}_{\rm ret}=\frac{{\bf r}\times{\bf E}_{\rm ret}}{r}\,,
\label
{H-L-W}
\end{equation}                       
where ${\bf r}$ is the radius vector drawn from the point of emission 
${\bf z}(t_{\rm ret})$ to the point of observation ${\bf x}$ in a particular Lorentz frame.

Let a particle be moving along a straight line $z^\mu(s)=z^\mu(0)+V^\mu s$, 
$V^\mu=$ const.
Then, in a Lorentz frame in which the time axis is parallel to $V^\mu$, we have 
$U^\mu=V^\mu$, and Eqs.~(\ref{F-L-W})--(\ref{V}) describe the Coulomb field.
In a sense this feature remains valid for the field generated by 
an arbitrarily moving charge.
Indeed, substituting Eqs.~(\ref{F-L-W})--(\ref{V}) into the expressions for the
electromagnetic field invariants 
\begin{equation}
{\cal S}=\frac12\,F_{\mu\nu}F^{\mu\nu}\,,
\label
{S-em-invariant}
\end{equation}                        
\begin{equation}
{\cal P}=\frac12\,F_{\mu\nu}{}^\ast\!F^{\mu\nu}\,
\label
{P-em-invariant}
\end{equation}                        
gives 
\begin{equation}
{\cal S}_{\rm ret}=-\frac{q^2}{\rho^4}\,,
\quad
{\cal P}_{\rm ret}=0\,.
\label
{em-invariant}
\end{equation}                        
Since ${\cal S}={\bf B}^2-{\bf E}^2$, and ${\cal P}=-2{\bf E}\cdot{\bf B}$, 
this result implies that, whatever smooth world line is chosen, 
one can find such a frame of reference  (which is peculiar to every point $x^\mu$)
that ${\bf B}_{\rm ret}=0$ and $\vert{\bf E}_{\rm ret}\vert=q/r^2$ in all points of
spacetime, that is, only electric field is observed in this frame.
Thus, there exists a global (noninertial) frame of reference in which the retarded 
electromagnetic field generated by a single arbitrarily
moving charge, shown in Eqs.~(\ref{F-L-W})--(\ref{V}), appears as a pure Coulomb 
field at each observation point.

The Li\'enard--Wiechert field (\ref{F-L-W}) is determined not only by the field 
$F_{\rm ret}$ as such but also by the frame of reference in which $F_{\rm ret}$ 
is measured. 
If we are to identify the net degrees of freedom related to $F_{\rm ret}$, 
irrespective of the used frame of reference, we conclude that just the Coulomb 
field is responsible for those degrees of freedom. 

This inference may seem surprising: any charged particle generates a field of  
electric type! 
However, we are well aware of the
fact that not only fields of electric type but also fields of magnetic type are available in
nature. 
Where do they come from? 
One can indicate at least two origins of  magnetic fields.
First, the superposition principle. 
It is an easy matter to verify from Eqs.~(\ref{F-L-W})--(\ref{V}) that the relations ${\cal P}=0$, ${\cal S}<0$ are 
in general no longer valid for electromagnetic fields generated by two or 
several charges, so that 
the configurations involved may well represent fields of magnetic type.
Note that the occurrence of pure
magnetic fields due to the circuition of electrons around closed paths suggests
a neutral system where electric fields of moving electrons and immovable nuclei
mutually cancel. 
Second, a pure magnetic field may also be related to spin and its
associated magnetic dipole moment of the charged particle, which, however, is
beyond the scope of the present discussion.

The general solution of the homogeneous wave equation (\ref{wave-A}) can be 
written as an arbitrary superposition 
of plane waves $\epsilon_\mu\exp(i k\cdot x)$ with a lightlike propagation
vector, $k^\mu$, $k^2=0$,
and the polarization vector $\epsilon_\mu$  
orthogonal to the propagation
vector, 
$\epsilon\cdot k=0$.
Since our interest here is only with fields distributed over empty space, it
is adequate to use a Fourier-integral expansion, so that the desired solution is  
\begin{equation}
{A}^{\mu}(x)=\int \alpha(k)\,\epsilon^\mu(k)\,e^{ik\cdot x}\,d^4k\,.
\label
{waves-}
\end{equation}  

We thus have established our assertion concerning the existence of two fundamental
solutions to Maxwell's equations giving rise to the variety of 
field configurations in classical electrodynamics, with due reservation of course
that all the field configurations in macroscopic media were left aside in
the present consideration.

\section{Discussion and outlook (for the expert reader)}
\label
{concluding}
Since the discovery of the Aharonov--Bohm effect\footnote{This effect would be 
more properly termed the Ehrenberg--Siday--Aharonov--Bohm effect because it was 
discovered by Ehrenberg and Siday 10 years before Aharonov and Bohm \cite{Ehrenberg, 
Aharonov}.}, the quantity $A_{\mu}$ achieved settled status of the 
basic variable for accounting of the electromagnetic field 
in quantum theory\footnote{Note, however, that this status of vector potentials was 
challenged in \cite{Vaidman}.
We will not go into details of this controversial issue, and refer the
interested reader to the original literature. }. 
Most current theories in high energy physics and gravity begin with gauge invariance  
as a first principle, that is, proceeding from vector 
potentials as the basic field variables.
To illustrate, we take a glance at the Yang--Mills--Wong theory \cite{YM, Wong}. 
The dynamical equations governing the  Yang--Mills field read
\begin{equation}
\partial_\mu G_a^{\mu\nu}+gf_{abc}A^b_\mu G^c_{\mu\nu}=4\pi j_a^\nu\,, 
\label
{YM-eq}
\end{equation}                                          
where $G_a^{\mu\nu}$ is the non-Abelian field strength which is expressed in terms of 
vector potentials $A^b_\mu$ as 
\begin{equation}
G_a^{\mu\nu}=\partial^\mu A_a^{\nu}-\partial^\nu A_a^{\mu}
+ gf_{a}^{~bc} A_b^{\mu}\, A_c^{\nu}\,,
\label
{F-in-terms-A-YMW}
\end{equation}                                          
$g$ is the Yang--Mills coupling constant, $f_{abc}$ is the structure constants
of the gauge group involved, and $j_a^\mu$ is the color charge current 
of $N$ point particles, each carrying the color charge $Q_I^a$, 
analogous to the current of $N$ electrically charged point particles,
\begin{equation}
j_a^\mu(x)=\sum_{I=1}^N\int_{-\infty}^\infty ds_I\,Q_{aI}(s_I)\,v_I^\mu(s_I)\,\delta^4\left[x-z_I(s_I)\right].
\label
{j-mu-YMW}
\end{equation}                                          

Let $T_a$ be the generators of the gauge group.
All color variables (the field strength, vector potentials, color charges, etc.)  
can be written in matrix notation, as exemplified by $G_{\mu\nu}=\frac{i}{g}T_aG^a_{\mu\nu}$.
Then equations (\ref{YM-eq}) and (\ref{F-in-terms-A-YMW}) become  
\begin{equation}
[D_\mu, G^{\mu\nu}]=4\pi  j^\nu\,, 
\label
{YM-eq-mtr}
\end{equation}                                          
\begin{equation}
G_{\mu\nu}=\partial_\mu A_{\nu}-\partial_\nu A_{\mu}
+ig\left[A_{\mu},\, A_{\nu}\right].
\label
{F-in-terms-A-YMW-matr}
\end{equation}                                          
Here, the square brackets stand for commutators of matrix-valued quantities, 
and $D_\mu$ is the so-called covariant Yang--Mills derivative whose action on 
any field $\phi=\phi^a T_a$, transforming 
according to the adjoint representation of the gauge group, is given by
\begin{equation}
D_\mu \phi=\partial_\mu \phi+g\,[A_\mu,\phi]\,.
\label
{covar-d-YM}
\end{equation}                   
Note that for any gauge covariant quantity $\phi$, 
\begin{equation}
[D_\mu,D_\nu]\,\phi=g\,[G_{\mu\nu},\phi]\,.
\label
{D-mu,D-nu=F-mu-nu}
\end{equation}
By recognizing that $G_{\mu\nu}$ is expressed in terms of 
$A_{\mu}$, according to (\ref{F-in-terms-A-YMW-matr}), we come to a 
condition underlying 
this relation, the Bianchi identity, 
\begin{equation}
[D_\lambda, G_{\mu\nu}]+[D_\nu, G_{\lambda\mu}]+[D_\mu, G_{\nu\lambda}]
=0\,, 
\label
{Bianchi-YM-eq-mtr}
\end{equation}                                          
which can be verified through the use of the Jacobi identity 
\begin{equation}
[D_\lambda,[D_{\mu},D_{\nu}]]+[D_\nu,[D_{\lambda},D_{\mu}]]+[D_\mu,[D_{\nu}, 
D_{\lambda}]]
=0\,,
\label
{Jacobi-Bianchi-YM}
\end{equation}                                          
combined with Eq.~(\ref{D-mu,D-nu=F-mu-nu}).

Of course if we define the field strength $G^a_{\mu\nu}$ in terms of vector 
potentials $A_\mu^a$ according to  (\ref{F-in-terms-A-YMW-matr}), then there is no need to 
join the Bianchi identity (\ref{Bianchi-YM-eq-mtr}) to the field equation 
(\ref{YM-eq-mtr}) for completing the dynamics of this theory. 
Since the gauge-dependent quantities $A^a_\mu$ appear in the Yang--Mills theory from 
the outset, the set of dynamical equations is no longer overdetermined.
The situation with the Bianchi identity in General Relativity closely resembles that in the Yang--Mills 
theory \cite{Misner}.

We thus see that classical electrodynamics offers a very instructive example of how the concept of 
gauge fields and gauge invariance can be introduced in their simplest 
physical and mathematical context.

One further feature of classical electrodynamics is its remarkably transparent 
field configuration arrangement: every field configuration stems from the 
Coulomb fields and plane waves.
Note, however, that our concern here is with the fundamental aspects, rather 
than practical uses, of this arrangement.
The Fourier-integral expansion (\ref{waves-}) is sound but not universally 
convenient.  
In some instances it would be appropriate to expand the solution of the 
homogeneous wave equation (\ref{wave-A}) in terms of spherical harmonics.
For example, a plausible guess about the nature of ball lighting is that the 
essential prerequisite to ball lighting formation is a 
steady-state field generated by converging and diverging axially symmetric 
microwaves \cite{Kapitza, Stenhoff}.
Self-dual solution to free Maxwell's equations \cite{cheon} may be suitable to 
the analysis of standing wave configurations in this as yet unsolved problem.

By contrast, exact solutions in the classical Yang--Mills theory and 
General Relativity pose many problems.
The dynamical equations of these theories are nonlinear, so that any 
superposition of solutions appears to be something other than a new solution.
Solutions to the classical Yang--Mills equations, known by the end of the 
1970s, are reviewed in Ref.~\cite{Actor}. 
Exact solutions of {quantum} Yang--Mills theory are altogether out of the
question. 
This task is among one of the seven problems  recorded by 
the Clay Mathematics Institute as  the Millennium Prize Problems--the most
difficult issues with which mathematicians were struggling at the turn of the
second millennium \cite{JaffeWitten}.
A large body of exact solutions in General Relativity are systematized in 
the catalog \cite{Kramer}.

Are there exact solutions of these theories similar to the  Li\'enard--Wiechert 
solution (\ref{V})?
Such solutions to the Yang--Mills equations were indeed found in Ref.~\cite{k91}
(for a review see Ref.~\cite{k98}),
and rediscovered in Ref.~\cite{Sarioglu}.
These solutions fall into two classes.
One of them contains solutions describing  Yang--Mills fields of electric
type, whose field invariants ${\cal P}$ and ${\cal S}$ built out of 
$G^a_{\mu\nu}$ are ${\cal P}=0$ and ${\cal S}<0$, while the
other contains solutions describing fields of magnetic type specified by
${\cal P}=0$ and ${\cal S}>0$.
These two types of solutions are very likely related to two phases of the 
Yang--Mills vacuum. 

As to General Relativity, exact solutions describing the gravitational 
field generated by an arbitrarily moving massive particle, similar to the 
Li\'enard--Wiechert solution, still remain unknown. 

The non-Abelian analogues of electromagnetic plane waves, that is, exact solutions
to equation (\ref{YM-eq-mtr}) with $j^\mu=0$ obeying the requirements that
the energy density is bounded throughout spacetime, the direction of the
Poynting vector is constant, and the magnitude of the Poynting vector is equal to 
the energy density, were found in Ref.~\cite{Coleman}.
However, such waves moving in different directions can not be superposed, and
hence, they are of no practical importance.

\end{document}